\documentclass[12pt]{article}

\begin{document}

\begin{titlepage}

\begin{flushright}
\end{flushright}
\vskip 2.5cm

\begin{center}
{\Large \bf Astrophysical Limits on Lorentz Violation for Pions}
\end{center}

\vspace{1ex}

\begin{center}
{\large Brett Altschul\footnote{{\tt baltschu@physics.sc.edu}}}

\vspace{5mm}
{\sl Department of Physics and Astronomy} \\
{\sl University of South Carolina} \\
{\sl Columbia, SC 29208 USA} \\

\end{center}

\vspace{2.5ex}

\medskip

\centerline {\bf Abstract}

\bigskip

Pions, like nucleons, are composed primarily of up and down quarks and gluons.
Constraints on spin-independent Lorentz violation in the proton, neutron, and pion
sectors translate into bounds on Lorentz violation
for the fundamental fields. The best bounds on pion
Lorentz violation come from astrophysical measurements. The absence of the
absorption process $\gamma\rightarrow\pi^{+}+\pi^{-}$ for up to 50 TeV photons
constrains the
possibility that pions' maximum achievable velocities are less than 1 at the
$1.5\times10^{-11}$ level. The fact that pions with energies up to 30 TeV are
observed to decay into photons rather than hadrons bounds the possibility of a
maximum velocity greater than 1 at the $2\times10^{-9}$ level. This provides
the first two-sided bounds on Lorentz violation for pions.

\bigskip

\end{titlepage}

\newpage

In the last ten years, there has been growing interest in the suggestion that
Lorentz symmetry may not be exact~\cite{ref-reviews}. In this scenario, special
relativity would be an excellent approximation to reality, but there would also exist
real deviations from its predictions. This kind of
Lorentz violation cannot occur in the standard model or any of its usual
generalizations. This means that the experimental discovery of Lorentz violation
would be a sure sign of significantly new physics. Thus far, there is no experimental
evidence for Lorentz violation, but new experiments and reanalyses of older data have
yielded some impressive bounds on many of the various coefficients that parameterize
Lorentz violation in effective field theory.

The effective field theory that describes Lorentz violation is the standard model
extension (SME)~\cite{ref-kost2}. In addition to the experimental
constraints that have been placed on many of the SME parameters, there has been quite
a bit of theoretical work exploring the structure of this effective theory.
Radiative corrections in the SME have curious
properties~\cite{ref-jackiw1,ref-altschul1}, and many novel
physical phenomena become possible when Lorentz symmetry is broken. By
studying these new phenomena, it is possible to place new constraints on the SME
coupling constants. Because Lorentz violation is known to be a small effect,
most analyses are performed only to leading order in the SME coefficients.

We will focus here on Lorentz violation in the pion sector. One reason for this
focus is
simply practical; there are bounds on pion Lorentz violation that are relatively easy
to establish once the theoretical groundwork is done. However, there is another
deeper reason why constraining the SME parameters for pions is interesting.
Like the proton and neutron (which together form the nucleon isodoublet),
the pion is composed primarily of up,
down, and gluon fields. To constrain the simplest (i.e. spin-independent) forms
of Lorentz violation for each of these three fundamental fields, we need bounds
for three different particle species. For fermion fields, the spin-independent
Lorentz violations are parameterized by coefficients known as $c^{\nu\mu}$,
for which there are very tight bounds in both the proton and neutron
sectors~\cite{ref-bear,ref-wolf}.
These translate into constraints on two particular linear combinations of the up, down,
and gluon coefficients. One of these combinations depends on
the detailed structure of the nucleon. It is a linear combination of the coefficients
for the up and down quarks---$c^{\nu\mu}_{u}$ and $c^{\nu\mu}_{d}$---and nine of the
nineteen gluon coefficients $k_{G}^{\mu\nu\rho\sigma}$. The relevant nine are the
gluon analogues of the $\tilde{\kappa}_{e-}$, $\tilde{\kappa}_{o+}$, and
$\tilde{\kappa}_{{\rm tr}}$ in the photon sector~\cite{ref-kost16}.
However, the second combination bounded by the proton and neutron experiments is,
to very good accuracy, the difference $c^{\nu\mu}_{u}-c^{\nu\mu}_{d}$.
Since this difference is very small, we know that the Lorentz violation respects
isospin symmetry to high accuracy.

The free Lagrange density we shall consider for the pion field is
\begin{equation}
{\cal L}=\frac{1}{2}(\partial^{\mu}\pi^{i})(\partial_{\mu}\pi^{i})+\frac{1}{2}
k^{\mu\nu}(\partial_{\mu}\pi^{i})(\partial_{\nu}\pi^{i})
-\frac{m^{2}}{2}\pi^{i}\pi^{i}.
\end{equation}
The pion field $\pi$ is an isotriplet
indexed by $i$. The charged pions are represented by
$\pi^{\pm}=\frac{1}{\sqrt{2}}(\pi^{1}\pm i\pi^{2})$, and $\pi^{3}$ is the field of
the neutral $\pi^{0}$.
${\cal L}$ contains an isospin-symmetric Lorentz-violating term, parameterized by
the nine-component, traceless symmetric tensor $k$. $k$ is another unknown linear
combination of $c_{u}$, $c_{d}$, and $k_{G}$. Any other
renormalizable SME terms would violate isospin symmetry.
(The number of Lorentz-violating operators that can be constructed with a
pseudoscalar field is rather limited.)
There is another possible renormalizable operator (an $a$ term) that involves
only the charged pion fields, but it is of diminishing importance at higher
energies.
We have neglected the (isospin-violating) mass difference between the
charged and neutral pions, purely for simplicity; the 3 percent
mass difference is not meaningful at the level of accuracy of our calculations.
Because $c^{\nu\mu}_{u}-c^{\nu\mu}_{d}$ is known to be extremely small
(effectively zero compared with the size of Lorentz violations we shall consider
here), any
isospin violations associated with
the Lorentz violation must be related to the very slight
differences in the structure of the charged and neutral pions caused by unequal up
and down masses and charges. This effect should be suppressed, just as is the mass
difference between the pion types, and can be similarly neglected.

The pions described by ${\cal L}$ have maximum achievable velocities (MAVs) that are
generally direction dependent. In the direction $\hat{e}$, the MAV
is~\cite{ref-altschul4}
\begin{equation}
v_{\max}\equiv1+\delta(\hat{e})=1-\frac{1}{2}\left[
k_{jk}\hat{e}_{j}\hat{e}_{k}+2k_{j0}\hat{e}_{j}+k_{00}\right],
\end{equation}
and at high energies the pion dispersion relation is effectively
\begin{equation}
E=\sqrt{m^{2}+[1+2\delta(\hat{p})]\vec{p}\,^{2}}.
\end{equation}

Electromagnetic interactions are, of course, not invariant under isospin. (We shall
assume a conventional photon sector, which is also well supported
experimentally~\cite{ref-carroll2,ref-kost11,ref-antonini,ref-stanwix,ref-herrmann,ref-carone}.)
The
charged pions are minimally coupled to the radiation field. Neutral pions decay
electromagnetically into two photons, primarily via the chiral anomaly. It is by
tracking the effects of these electromagnetic interactions that we can place bounds
on $k$. Bounding the pion $k$ will constrain the remaining unbounded
linear combination of spin-independent up, down, and gluon coefficients.

Since the most energetic particles we can study are astrophysical, the best
bounds on many SME coefficients come from analyses of high-energy
astrophysical processes~\cite{ref-jacobson1,ref-altschul6,ref-altschul15}.
Previously, we placed bounds on the $k$ coefficients by considering
processes involving charged pions, as
part of an analysis that applied to all charged particles~\cite{ref-altschul14}.
If $k$ is such that the MAV for a charged species $X$ is less than 1 in a given
direction, the reaction $\gamma\rightarrow X^{+}+X^{-}$ is allowed if the
photon energy is large enough~\cite{ref-stecker}. This process (which is a form of
vacuum photon
absorption) would prevent sufficiently energetic photons from reaching Earth. The
fact that TeV photons can reach Earth from virtually any direction places bounds on
the appropriate Lorentz violation coefficients at the $10^{-15}(m_{X}^{2}/m_{e}^{2})$
level. For the charged pions, this means bounds on $k$ of $10^{-10}$--$10^{-11}$.
However, to get a two-sided bound on $k$, we must constrain the possibility that
the MAV
may be greater than 1 as well. To find such a bound, we must look to the neutral
pions and understand how Lorentz violation affects their decays.

The dominant decay mode for neutral pions is $\pi^{0}\rightarrow2\gamma$, and
the lifetime of a $\pi^{0}$ at rest is less than $10^{-16}$ s.
However, there can be
interesting changes in how neutral pions decay if Lorentz violation is present.
For pions with large momenta, new decay modes may be possible.
In particular, if the pion energy exceeds $E_{N}=\sqrt{(4M_{N}^{2}-m^{2})/2
\delta(\hat{p})}$, where $M_{N}$ is the nucleon mass,
the process $\pi^{0}\rightarrow N+\bar{N}$ becomes allowed. (To the level of accuracy
we will be interested in, the nucleon sector is free of Lorentz violation.)
Since the $\pi^{0}$ couples very strongly to nucleons, this decay proceeds much more
quickly than the electromagnetic one, and above this threshold, the pions will decay
into nucleons rather than photons.
Because of ultrarelativistic beaming effects,
the particles produced in a $\pi^{0}$ decay travel in essentially the same direction
$\hat{e}$ as their progenitor. So if a pion is present in an energetic source, it
must be moving
in the source-to-Earth direction for its decay products to be detected by our
telescopes.
This means that the
observation on Earth of pion decay photons of energy $E$ propagating in a direction
$\hat{e}$ sets a bound on $k$ that can be parameterized as
\begin{equation}
\label{eq-deltabound1}
\delta(\hat{e})<\frac{2M_{N}^{2}}{E^{2}}.
\end{equation}
This is the main effect that we shall use to place bounds on the possibility of
a positive $\delta(\hat{e})$, although it is not the only potentially interesting
modification of $\pi^{0}$ decay.

When particles possess
Lorentz-violating dispersion relations, there may also be upper thresholds for
certain decays. These are energies above which a particle 
cannot decay, even though it might decay at rest.
Naively, we might expect this kind
of threshold to be associated with dispersions relations that allow superluminal
speeds. If $v>1$, the time dilation factor would seem to diverge, allowing the
particle to live for an infinitely long time. However, this intuitive argument turns
out not to be correct.

In the right parameter regime, there is actually an
upper threshold for $\pi^{0}\rightarrow2\gamma$.
If the initial $\pi^{0}$ has momentum $\vec{p}$, the decay photons must together have
an energy of at least $\left|\vec{p}\,\right|$. If the energy $E$ of the pion is
less than this, the decay cannot proceed, and this sets an upper threshold at
$E_{\gamma}=m/\sqrt{-2\delta(\hat{p})}$. Note that this threshold only exists if
$\delta(\hat{p})<0$, corresponding to a MAV smaller than 1.

The fact that $\pi^{0}\rightarrow2\gamma$ may be forbidden when the MAV is less than
1 suggests that it might be possible to place two-sided bounds on $k$ by looking
only at neutral pions (i.e. without making use of isospin invariance). Yet while this
is true in theory, there is a significant difficulty.
That the process $\pi^{0}\rightarrow2\gamma$ is forbidden does not mean
that sufficiently energetic pions cannot transfer their energies into
$\gamma$-rays, because the
usual decay mode is not the only possible process involving a pion and two photons.
If $\delta(\hat{p})
<0$, the pion energy grows more slowly as a function of $\left|\vec{p}\,\right|$ than
is usual. What prevents the two-photon decay at high energies is that the pion may
have insufficient energy to produce the photons. However, another process
replaces it above the threshold. The
reaction $\gamma+\pi^{0}\rightarrow\gamma$ (``absorption'' of a pion by a photon)
is normally forbidden by energy-momentum
conservation; in the pion's rest frame, the pion contributes energy but not momentum
to the system, but the final photon cannot carry away this excess energy. However, if
the pion has less energy as a function of momentum than $\sqrt{m^{2}+\vec{p}\,^{2}}$,
the process can occur. In fact, at exactly $E_{\gamma}$, it becomes kinematically
allowed. The matrix element for the new process is of the same order as for
the conventional two-photon decay.

The new process does require a minimum initial photon energy to occur. In the
threshold 
configuration, the pion and photon collide head on. The energy the incoming photon
must possess in order to trigger a reaction with a pion of momentum $\vec{p}$ is
\begin{equation}
\label{eq-epsilon}
\epsilon_{\gamma}=-\delta(\hat{p})\left|\vec{p}\,\right|-\frac{m^{2}}
{2\left|\vec{p}\,\right|}.
\end{equation}
For pion energies $E>E_{\gamma}$, $\epsilon_{\gamma}$ grows as a function of
$\left|\vec{p}\,\right|$. Near threshold, the two terms in (\ref{eq-epsilon}) are
comparable in size.

The $\gamma$-ray signature of the $\delta(\hat{p})<0$ pion dispersion relation is
therefore quite
peculiar. At lower energies, two-photon decays are allowed. Higher-energy pions only
produce a single photon when
they disintegrate, so there is a diminution in the photon spectrum below
$E_{\gamma}$; just below this threshold, the photon production will fall almost to
zero. Plentiful production of photons with energies above $E_{\gamma}$ is
again possible, but only as long there are seed photons with energies above
$\epsilon_{\gamma}$. In practice, since $\epsilon_{\gamma}$ is suppressed relative to
the pion energy by an ${\cal O}(k)$ factor, it seems unlikely that an absence of
sufficiently energetic photons would cut off the spectrum. If there are extremely
energetic pions present in a source, there are also presumably going to be present
the much lower-energy photons needed for the $\gamma+\pi^{0}\rightarrow\gamma$
reaction.

So while we have identified a very peculiar feature of $\pi^{0}$
disintegration in the $\delta(\hat{p})<0$ regime, it is not so useful for setting
experimental bounds on $k$.
One might hope to distinguish between $\pi^{0}\rightarrow2\gamma$ and
$\gamma+\pi^{0}\rightarrow\gamma$ with careful modeling of sources, but that may not
be a realistic goal. Provided the seed photons needed to initiate
$\gamma+\pi^{0}\rightarrow\gamma$
are plentiful, this process will occur rapidly, and its rate will be
determined by the rate at which the energetic pions are produced in hadronic
collisions. Having pions disintegrate via this reaction merely cuts the
number of $\gamma$-rays produced in half while
roughly doubling their energies. Source models
are not typically accurate enough to distinguish between scenarios that differ only
by such factors of 2.

So the most reliable astrophysical techniques for placing bounds on pion Lorentz
violation are as follows. If pions moving in the direction $\hat{e}$ decay into
$\gamma$-rays, which are then observed on Earth, this constrains $\delta(\hat{e})$
according to (\ref{eq-deltabound1}).
The complementary bound is the charged pion one already discussed
in~\cite{ref-altschul14}. The arrival of any energetic photon along a direction
$\hat{e}$, sets the bound
\begin{equation}
\label{eq-deltabound2}
\delta(\hat{e})>-\frac{2m^{2}}{E^{2}},
\end{equation}
because otherwise, the photon would decay into a collinear $\pi^{+}\pi^{-}$ pair.

With data from enough sources across the sky, it is possible to constrain all nine
components of $k$ to a bounded region of parameter space. The trickiest part about
setting these bounds is identifying definitely hadronic sources of $\gamma$-rays.
In such sources, pions are generated in highly energetic proton-proton collisions,
and very-high-energy $\gamma$-rays are produced primarily through the
decay of the neutral pions. However, distinguishing these hadronic $\gamma$-rays
from ones produced by electrons
through inverse Compton scattering can be a challenge.
Fortunately, such electrons are also strong emitters of synchrotron x-rays, and the
x-ray spectrum combined with data on the structure of the source may be
inconsistent with significant inverse Compton production at the highest photon
energies.

Any source whose $\gamma$-ray spectrum can be identified as originating in
$\pi^{0}$ decay automatically provides a two-sided bound on the relevant
$\delta(\hat{e})$, since both (\ref{eq-deltabound1}) and (\ref{eq-deltabound2})
apply, although the bound coming from (\ref{eq-deltabound1}) is more than an order
of magnitude weaker because of the relative factor
$M_{N}^{2}/m^{2}$. $\gamma$-rays produced by processes other than $\pi^{0}$
decay also set bounds of the form (\ref{eq-deltabound2}).

\begin{table}
\begin{center}
\begin{tabular}{|l|c|c|c|c|c|}
\hline
Emission source & $\hat{e}_{X}$ & $\hat{e}_{Y}$ & $\hat{e}_{Z}$ & $E$ &
References \\
\hline
Berkeley 87 & $-0.46$ & 0.65 & $-0.61$ &
10 TeV & \cite{ref-abdo,ref-bednarek} \\
Cygnus OB2 & $-0.46$ & 0.59 & $-0.66$ &
10 TeV & \cite{ref-aharonian16,ref-bednarek} \\
RX J0852.0-4622 & 0.46 & $-0.51$ & 0.72 &
10 TeV & \cite{ref-aharonian12,ref-berezhko1} \\
RX J1713-3946 & 0.16 & 0.75 & 0.64 &
30 TeV & \cite{ref-aharonian13,ref-berezhko2,ref-cyhuang} \\
Sagittarius B2 & 0.06 & 0.88 & 0.47 &
15 TeV & \cite{ref-aharonian14,ref-crocker1} \\
Westerlund 2 & 0.49 & $-0.22$ & 0.84 &
15 TeV & \cite{ref-aharonian15,ref-bednarek} \\
\hline
\end{tabular}
\caption{
\label{table-pisources}
Parameters for astrophysical sources with hadronic $\gamma$-ray spectra. $E$ is
the maximum energy observed for $\pi^{0}$ disintegration photons.
The coordinates $X$, $Y$, and $Z$ are in sun-centered celestial equatorial
coordinates~\cite{ref-bluhm4}.
References are given for each source.}
\end{center}
\end{table}

Because of uneven sky coverage, the bounds on the individual $k$ coefficients are
relatively weak, although they can be extracted from the bounds of the form
(\ref{eq-deltabound1}) and (\ref{eq-deltabound2}) by linear programming. Sources
useful for setting photon absorption bounds were tabulated in~\cite{ref-altschul14}.
Table~\ref{table-pisources} gives a number of sources whose very-high-energy
$\gamma$-ray spectra have been identified as having hadronic origins. Such sources
are often associated with dense molecular clouds, which energetic protons interact
with to produce copious pions.

The tightest and most easily expressible constraints are possible if the pion MAV is
assumed to be independent of direction. Then there is a single parameter $\delta$,
such that $k_{00}=-\frac{3\delta}{2}$, $k_{j0}=0$, and $k_{jk}=-\frac{\delta}{2}
\delta_{jk}$. The highest-energy photons that are observed reaching the earth come
from the Crab nebula; Crab $\gamma$-rays have energies of up to at least 50
TeV~\cite{ref-tanimori},
and the absence of $\gamma\rightarrow\pi^{+}+\pi^{-}$ sets the
bound
\begin{equation}
\label{eq-chargedbound}
\delta>-1.5\times 10^{-11}.
\end{equation}
However, these $\gamma$-rays do not appear to originate in $\pi^{0}$ disintegration.
Pion decay photons reach Earth from the supernova remnant RX J1713-3946 with
energies of up to 30 TeV. That pions this energetic should decay into
photons rather than hadrons requires that
\begin{equation}
\delta<2\times 10^{-9}.
\end{equation}

If the TeV $\gamma$-rays from more sources can be identified as being hadronic in
origin, the bounds on the individual $k$ coefficients will improve, probably
to the $10^{-9}$ level when full sky coverage is reached. However,
this may be difficult with the current generation of $\gamma$-ray telescopes.
Complementary bounds on the coefficients from different analyses could
improve the results as well; presently, the $\gamma$-ray data provide the
only constraints in the pion sector.

In summary, Lorentz violation in the pion sector is interesting both theoretically
and experimentally.
For a pseudoscalar particle like the pion, there are relatively few
possible renormalizable, Lorentz-violating operators. Isospin symmetry
restricts the operators even further, to the traceless symmetric tensor
$k$, which controls a pion's MAV. Bounds on $k$, in conjunction with tighter
bounds for the proton and neutron coefficients, translate into comparable bounds on
elements of the gluon tensor $k_{G}$ and the quark $c_{u}$ and $c_{d}$ separately.
Every astrophysical
$\gamma$-ray source provides a one-sided bound on the pion MAV in the
source-to-Earth direction. For sources where the origin of the most energetic
$\gamma$-rays can be traced to neutral pion decay, there is a two-sided bound on
the MAV. If spatial isotropy is assumed, the bounds are at the $10^{9}$--$10^{-11}$
level. These are the first two-sided bounds on Lorentz violation for pions.

%
%

\section*{Acknowledgments}
The author is grateful to V. A. Kosteleck\'{y} for helpful comments.


\begin{thebibliography}{99}

\bibitem{ref-reviews}For overviews of recent work on Lorentz violation, see D.
Mattingly, Living Rev. Rel. {\bf 5}, (2005) and the contents of {\em Proceedings of
the Third Meeting on {\rm CPT} and Lorentz Symmetry}, edited by V. A. Kosteleck\'{y}
(World Scientific, Singapore, 2005).
\bibitem{ref-kost2}D. Colladay, V. A. Kosteleck\'{y}, Phys. Rev. D {\bf 58},
116002 (1998).
\bibitem{ref-jackiw1}R. Jackiw, V. A. Kosteleck\'{y}, Phys. Rev. Lett.
{\bf 82}, 3572 (1999).
\bibitem{ref-altschul1}B. Altschul, Phys. Rev. D {\bf 69}, 125009 (2004).
\bibitem{ref-bear}D. Bear, R. E. Stoner, R. L. Walsworth, V. A. Kosteleck\'{y},
C. D. Lane, Phys. Rev. Lett. {\bf 85}, 5038 (2000).
\bibitem{ref-wolf}P. Wolf, F. Chapelet, S. Bize, A. Clairon,  Phys. Rev. Lett.
{\bf 96}, 060801 (2006).
\bibitem{ref-kost16}V. A. Kosteleck\'{y}, M. Mewes, Phys. Rev. D {\bf 66}, 056005
(2002).
\bibitem{ref-altschul4}B. Altschul, D. Colladay, Phys. Rev. D {\bf 71}, 125015
(2005).
\bibitem{ref-carroll2}S. M. Carroll, G. B. Field, Phys. Rev. Lett. {\bf 79},
2394 (1997).
\bibitem{ref-kost11}V. A. Kosteleck\'{y}, M. Mewes, Phys. Rev. Lett. {\bf 87},
251304 (2001).
\bibitem{ref-antonini}P. Antonini, M. Okhapkin, E. G\"{o}kl\"{u}, S. Schiller, Phys.
Rev. A {\bf 71}, 050101 (R) (2005).
\bibitem{ref-stanwix}P. L. Stanwix, M. E. Tobar, P. Wolf, M. Susli, C. R. Locke,
E. N. Ivanov, J. Winterflood, F. van Kann, Phys. Rev. Lett. {\bf 95}, 040404
(2005).
\bibitem{ref-herrmann}S. Herrmann, A. Senger, E. Kovalchuk, H. M\"{u}ller, A. Peters,
Phys. Rev. Lett. {\bf 95}, 150401 (2005).
\bibitem{ref-carone}C. D. Carone, M. Sher, M. Vanderhaeghen, Phys. Rev. D {\bf 74},
077901 (2006).
\bibitem{ref-jacobson1}T. Jacobson, S. Liberati, D. Mattingly, Nature {\bf 424},
1019 (2003).
\bibitem{ref-altschul6}B. Altschul, Phys. Rev. Lett. {\bf 96}, 201101 (2006).
\bibitem{ref-altschul15}B. Altschul, Phys. Rev. D {\bf 75}, 041301 (R) (2007).
\bibitem{ref-altschul14}B. Altschul, Astropart. Phys. {\bf 28}, 380 (2007).
\bibitem{ref-stecker}F. W.  Stecker, S. L.  Glashow, Astropart. Phys. {\bf 16},  97
(2001).

\bibitem{ref-abdo}A. A. Abdo, {\em et al.}, Astrophys. J. {\bf 658}, L33 (2007).
\bibitem{ref-bednarek}W. Bednarek, Mon. Not. R, Astron. Soc. {\bf 382}, 367 (2007).
\bibitem{ref-aharonian16}F. Aharonian, {\em et al.}, Astron. Astrophys. {\bf 393}, L37 (2002).
\bibitem{ref-aharonian12}F. A. Aharonian, {\em et al.}, Astron. Astrophys. {\bf 437},
L7 (2005).
\bibitem{ref-berezhko1}E. G. Berezhko, G. P\"{u}hlhofer, H. J. V\"{o}lk,
arxiv:0707.4661.
\bibitem{ref-aharonian13}F. A. Aharonian, {\em et al.}, Astron. Astrophys. {\bf 449},
223 (2006).
\bibitem{ref-berezhko2}E. G. Berezhko, H. J. V\"{o}lk, Astron. Astrophys. {\bf 451}, 981 (2006).
\bibitem{ref-cyhuang}C.-Y. Huang, S.-E. Park, M. Pohl, C. D. Daniels, Astropart.
Phys. {\bf 27}, 429 (2007).
\bibitem{ref-aharonian14}F. A. Aharonian, {\em et al.}, Nature {\bf 439}, 695 (2006).
\bibitem{ref-crocker1}R. M. Crocker, D. Jones, R. J. Protheroe, J. Ott, R. Ekers, F.
Melia, T. Stanev, A. Green, Astrophys. J. {\bf 666}, 934 (2007).
\bibitem{ref-aharonian15}F. Aharonian, {\em et al.}, Astron. Astrophys. {\bf 467}, 1075 (2007).

\bibitem{ref-bluhm4}R. Bluhm, V. A. Kosteleck\'{y}, C. D. Lane, N. Russell, Phys.
Rev. D {\bf 68}, 125008 (2003).
\bibitem{ref-tanimori}T. Tanimori, {\em et al.}, Astrophys. J. {\bf 492}, L33 (1998).

\end{thebibliography}
\end{document}